\documentstyle[prd,aps,preprint,tighten,epsfig]{revtex}

\begin{document}

\draft

\preprint{\begin{tabular}{r}
\framebox{\bf hep-ph/0107123} \\
{~}
\end{tabular}}

\title{More Straightforward Extraction of the Fundamental Lepton
Mixing Parameters from Long-Baseline Neutrino Oscillations}
\author{\bf Zhi-zhong Xing}
\address{Institute of High Energy Physics, P.O. Box 918 (4),
Beijing 100039, China \\
({\it Electronic address: xingzz@mail.ihep.ac.cn}) }
\maketitle

\begin{abstract}
We point out the simple reversibility between the fundamental neutrino mixing
parameters in vacuum and their effective counterparts in matter.
The former can therefore be expressed in terms of the latter, allowing more
straightforward extraction of the genuine lepton mixing quantities
from a variety of long-baseline neutrino oscillation experiments.
In addition to the parametrization-independent results,
we present the formulas based on the standard parametrization
of the lepton flavor mixing matrix and give a typical numerical
illustration.
\end{abstract}


\newpage

\section{Introduction}

The observation of solar and atmospheric neutrino oscillations
in the Super-Kamiokande experiment has provided the most convincing
evidence that neutrinos are massive and lepton flavors are mixed \cite{SK}
\footnote{The latest SNO measurement \cite{SNO},
together with the Super-Kamiokande data, has provided the first direct
evidence that there is a muon- and (or) tau-neutrino component in the
solar electron-neutrino flux.}.
It opens an important window for deep insight into the dynamics of
fermion mass generation and flavor mixing in particle physics,
and has important implications in astrophysics and cosmology.

Although some useful constraints on the lepton flavor mixing matrix
can be obtained from the Super-Kamiokande measurements and other
non-accelerator neutrino experiments,
a precise determination of its parameters has to rely on the
new generation of accelerator neutrino experiments with very long
baselines \cite{Long}, including
possible neutrino factories \cite{Factory}. The terrestrial matter
effects in all long-baseline neutrino experiments
must be taken into acount, since they unavoidably modify the genuine
behaviors of neutrino oscillations in vacuum.

To express the pattern of neutrino oscillations in matter in
the same form as that in vacuum, one may define the {\it effective}
neutrino masses $\tilde{m}_i$ and the {\it effective} lepton flavor
mixing matrix $\tilde{V}$ in which the terrestrial matter effects
are already included.
In this commonly-accepted approach, it is necessary to find out the
concrete relations between the fundamental quantities of
lepton mixing in vacuum ($m_i$ and $V$)
and their effective counterparts in matter ($\tilde{m}_i$ and $\tilde{V}$).
The analytically exact formulas of $\tilde{m}_i$ and $\tilde{V}$
as functions of $m_i$ and $V$ have been achieved in the three-neutrino
mixing scheme \cite{Barger,Zaglauer,Xing00}, but no effort has
been made to derive the expressions of $m_i$ and $V$ in terms of
$\tilde{m}_i$ and $\tilde{V}$. The latter case is equivalently
interesting in phenomenology, because our physical purpose is
to determine the fundamental parameters of lepton flavor mixing
from the effective ones, which can directly be observed from a variety
of long-baseline neutrino oscillation experiments.

This paper aims to present the analytically exact results of
the fundamental quantities $m_i$ and $V$ in terms of the effective
quantities $\tilde{m}_i$ and $\tilde{V}$. In Section 2, we first point
out the interesting reversibility between $(m_i, V)$ and
$(\tilde{m}_i, \tilde{V})$, and then derive their explicit relations
in a way independent of any specific
parametrizations of the lepton flavor mixing matrix.
In Section 3, the standard parametrization is introduced to describe
$V$ and $\tilde{V}$ in a parallel form. We calculate the mixing
angles and the CP-violating phase(s) of $V$ as functions of the
relevant parameters of $\tilde{V}$, and give a typical numerical
illustration. Section 4 is devoted to a brief summary.

\section{Reversibility and parametrization-independent formulas}

Let us concentrate on the flavor mixing between three active neutrinos
and three charged leptons.
The $3\times 3$ lepton flavor mixing matrix $V$ in vacuum
is defined to link the neutrino mass eigenstates
($\nu_1, \nu_2, \nu_3$) to the neutrino flavor eigenstates
($\nu_e, \nu_\mu, \nu_\tau$):
\begin{equation}
V \; =\; \left ( \matrix{
V_{e1}		& V_{e2}	& V_{e3} \cr
V_{\mu 1}	& V_{\mu 2}	& V_{\mu 3} \cr
V_{\tau 1}	& V_{\tau 2}	& V_{\tau 3} \cr} \right ) \; .
\end{equation}
A similar definition can be
made for $\tilde{V}$, the effective counterpart of $V$ in matter.
If neutrinos are massive Dirac fermions, $V$ or $\tilde{V}$
can be parametrized
in terms of three rotation angles and one CP-violating phase.
If neutrinos are Majorana fermions, however, two additional
CP-violating phases are in general needed to fully parametrize $V$
or $\tilde{V}$.
The strength of CP or T violation in neutrino oscillations, no
matter whether neutrinos are Dirac or Majorana particles,
depends only upon a universal parameter $J$ (in vacuum)
or $\tilde{J}$ (in matter) \cite{Jarlskog}:
\begin{eqnarray}
{\rm Im} \left (V_{\alpha i}V_{\beta j} V^*_{\alpha j}V^*_{\beta i} \right )
& = & J \sum_{\gamma,k} \left (\epsilon^{~}_{\alpha \beta \gamma}
\epsilon^{~}_{ijk} \right ) \; ,
\nonumber \\
{\rm Im} \left (\tilde{V}_{\alpha i}\tilde{V}_{\beta j}
\tilde{V}^*_{\alpha j}\tilde{V}^*_{\beta i} \right )
& = & \tilde{J} \sum_{\gamma,k} \left (\epsilon^{~}_{\alpha \beta \gamma}
\epsilon^{~}_{ijk} \right ) \; ,
\end{eqnarray}
where the Greek subscripts $(\alpha, \beta, \gamma)$ and the Latin subscripts
$(i, j, k)$ run over $(e, \mu, \tau)$ and $(1, 2, 3)$, respectively.
Of course, $J$ and $\tilde{J}$ are two rephasing-invariant quantities.

The effective Hamiltonian responsible for the propagation of neutrinos
in vacuum or in matter can be written as
\begin{eqnarray}
H_\nu & = & \frac{1}{2E} ~ V \left (\matrix{
m^2_1     & 0      & 0 \cr
0         & m^2_2  & 0 \cr
0         & 0      & m^2_3 \cr} \right ) V^{\dagger} \; ,
\nonumber \\
\tilde{H}_\nu & = & \frac{1}{2E} ~ \tilde{V} \left (\matrix{
\tilde{m}^2_1     & 0      & 0 \cr
0         & \tilde{m}^2_2  & 0 \cr
0         & 0      & \tilde{m}^2_3 \cr} \right ) \tilde{V}^{\dagger} \; ,
\end{eqnarray}
where $E$ is the neutrino beam energy;
$m_i$ and $\tilde{m}_i$ (for $i=1, 2, 3$) denote the
neutrino masses in vacuum and those in matter, respectively.
The deviation of $\tilde{H}_\nu$ from $H_\nu$ results
non-trivially from the
charged-current contribution to the $\nu_e e^-$ forward
scattering \cite{Wolfenstein},
when neutrinos travel through a normal material medium like the earth:
\begin{equation}
\Delta H_\nu  \; \equiv \; \tilde{H}_\nu - H_\nu =
\left ( \matrix{
a	& 0 	& 0 \cr
0	& 0	& 0 \cr
0	& 0	& 0 \cr } \right ) \; ,
\end{equation}
where $a = \sqrt{2} G_{\rm F} N_e$ with
$N_e$ being the background density of electrons.
Subsequently we assume a constant earth density profile
(i.e., $N_e$ = constant), which is a very good approximation
for all of the presently-proposed long-baseline neutrino experiments
\footnote{A careful analysis with the method of the Fourier series
expansion demonstrates that the matter profile effect is essentially
irrelevant, if the baseline of neutrino oscillations is about
3000 km or shorter \cite{Sato}. Therefore one may safely assume $N_e$
to be a constant in those realistic medium- and long-baseline neutrino
oscillation experiments with $L \leq 3500$ km. For $L >3500$ km
(in particular, $L \geq 5000$ km), however, details of the earth density
profile have to be taken into account towards an accurate description
of the features of terrestrial neutrino oscillations.}.

With the help of Eqs. (3) and (4), we obtain the following relations:
\begin{eqnarray}
\left ( \matrix{
m^2_1 	& 0	& 0 \cr
0	& m^2_2	& 0 \cr
0	& 0	& m^2_3 \cr } \right ) & = &
V^\dagger \left [ \tilde{V} \left ( \matrix{
\tilde{m}^2_1	& 0	& 0 \cr
0	& \tilde{m}^2_2	& 0 \cr
0	& 0	& \tilde{m}^2_3 \cr } \right )
\tilde{V}^\dagger - \left ( \matrix{
A	& 0 	& 0 \cr
0	& 0	& 0 \cr
0	& 0	& 0 \cr } \right ) \right ] V \; ,
\nonumber \\ \nonumber \\
\left ( \matrix{
\tilde{m}^2_1 	& 0	& 0 \cr
0	& \tilde{m}^2_2	& 0 \cr
0	& 0	& \tilde{m}^2_3 \cr } \right ) & = &
\tilde{V}^\dagger \left [ V \left ( \matrix{
m^2_1	& 0	& 0 \cr
0	& m^2_2	& 0 \cr
0	& 0	& m^2_3 \cr } \right )
V^\dagger + \left ( \matrix{
A	& 0 	& 0 \cr
0	& 0	& 0 \cr
0	& 0	& 0 \cr } \right ) \right ] \tilde{V} \; ,
\end{eqnarray}
where $A = 2aE$ depending linearly on the neutrino beam energy $E$.
This result implies that the expression of $\tilde{m}_i$ as
a function of $A$, $m_j$ and $V_{ej}$ is
reversible with that of $m_i$, which must behave via
the same function of variables $A$, $\tilde{m}_j$ and
$\tilde{V}_{ej}$; namely,
\begin{eqnarray}
m_i & = & f_i (\tilde{m}_j, ~ |\tilde{V}_{ej}|, ~ -A) \; ,
\nonumber \\
\tilde{m}_i & = & f_i (m_j, ~ |V_{ej}|, ~ +A) \; .
\end{eqnarray}
As $m_i$ is real and positive, it depends
only upon the absolute values of $\tilde{V}_{\alpha j}$.
The specific texture of $\Delta H_\nu$ assures that only
the matrix elements $|\tilde{V}_{e j}|$, which are always
associated with the matter parameter $A$, can affect $m_i$.
The similar argument is valid for $\tilde{m}_i$.
Analogously, there exists the simple reversibility between the
expressions of $V_{\alpha i}$ and $\tilde{V}_{\alpha i}$:
\begin{eqnarray}
V_{\alpha i} & = & F_{\alpha i} (\tilde{m}_j, ~ \tilde{V}_{\beta j}, ~ -A) \; ,
\nonumber \\
\tilde{V}_{\alpha i} & = & F_{\alpha i} (m_j, ~ V_{\beta j}, ~ +A) \; .
\end{eqnarray}
We observe that the reversible relations in Eqs. (6) and (7)
coincide formally with the sum rules of neutrino mass and mixing parameters
obtained in Ref. \cite{Xing01}:
\begin{equation}
\sum^3_{i=1} \left (m^2_i V_{\alpha i} V^*_{\beta i} \right )
\; =\; \sum^3_{i=1} \left (\tilde{m}^2_i \tilde{V}_{\alpha i}
\tilde{V}^*_{\beta i} \right ) \; ,
\end{equation}
where $\alpha \neq \beta$ running over $(e, \mu, \tau)$.
In the limit $A =0$, $\tilde{m}_i = m_i$ and
$\tilde{V}_{\alpha i} = V_{\alpha i}$ must hold. Hence the
functions $f_i$ and $F_{\alpha i}$ might have transparent dependence
on $A$. To find out the explicit forms of $f_i$ and $F_{\alpha i}$
is simply an algebraic exercise in the three-neutrino mixing
scheme \cite{Barger,Zaglauer,Xing00}.

Using Eqs. (6) and (7), one may obtain the formulas of
$m_i$ and $V_{\alpha i}$ straightforwardly from those of
$\tilde{m}_i$ and $\tilde{V}_{\alpha i}$, which have been
presented in Ref. \cite{Xing00} in a parametrization-independent way.
We then arrive at
\begin{eqnarray}
m^2_1 & = & \tilde{m}^2_1 + \frac{1}{3} \tilde{x} -
\frac{1}{3} \sqrt{\tilde{x}^2 - 3\tilde{y}}
\left [\tilde{z} + \sqrt{3 \left (1-\tilde{z}^2 \right )} \right ] \; ,
\nonumber \\
m^2_2 & = & \tilde{m}^2_1 + \frac{1}{3} \tilde{x} -
\frac{1}{3} \sqrt{\tilde{x}^2 -3\tilde{y}}
\left [\tilde{z} - \sqrt{3 \left (1-\tilde{z}^2 \right )} \right ] \; ,
\nonumber \\
m^2_3 & = & \tilde{m}^2_1 + \frac{1}{3} \tilde{x} +
\frac{2}{3} \tilde{z} \sqrt{\tilde{x}^2 - 3\tilde{y}} \;\; ,
\end{eqnarray}
where $\tilde x$, $\tilde y$ and $\tilde z$ are given by
\begin{eqnarray}
\tilde{x} & = & \Delta \tilde{m}^2_{21} + \Delta \tilde{m}^2_{31}
- A \; ,
\nonumber \\
\tilde{y} & = & \Delta \tilde{m}^2_{21} \Delta \tilde{m}^2_{31}
- A \left [
\Delta \tilde{m}^2_{21} \left ( 1 - |\tilde{V}_{e2}|^2 \right ) +
\Delta \tilde{m}^2_{31} \left ( 1 - |\tilde{V}_{e3}|^2 \right ) \right ] \; ,
\nonumber \\
\tilde{z} & = & \cos \left [ \frac{1}{3} \arccos
\frac{2\tilde{x}^3 -9\tilde{x}\tilde{y} - 27
A \Delta \tilde{m}^2_{21} \Delta \tilde{m}^2_{31}
|\tilde{V}_{e1}|^2}
{2 \left (\tilde{x}^2 - 3\tilde{y} \right )^{3/2}} \right ]
\end{eqnarray}
with $\Delta \tilde{m}^2_{21} \equiv \tilde{m}^2_2 - \tilde{m}^2_1$ and
$\Delta \tilde{m}^2_{31} \equiv \tilde{m}^2_3 - \tilde{m}^2_1$.
Furthermore, $V$ reads as follows:
\begin{equation}
V_{\alpha i} \; = \; \frac{\tilde{N}_i}{\tilde{D}_i}
\tilde{V}_{\alpha i} ~ - ~ \frac{A}{\tilde{D}_i}
\left [ \left (m^2_i - \tilde{m}^2_j \right )
\tilde{V}_{\alpha k} \tilde{V}^*_{e k}
+ \left (m^2_i - \tilde{m}^2_k \right )
\tilde{V}_{\alpha j} \tilde{V}^*_{e j} \right ] \tilde{V}_{e i} \; ,
\end{equation}
where $\alpha$ runs over $(e, \mu, \tau)$;
$i$, $j$ or $k$ runs over $(1, 2, 3)$ with $i \neq j \neq k$; and
\footnote{In obtaining Eqs. (11) and (12), we have required that
$\tilde{N}_i$ and $\tilde{D}_i$ have the same sign. Therefore
$A=0$ leads to $\tilde{N}_i/\tilde{D}_i =1$ and
$V_{\alpha i} = \tilde{V}_{\alpha i}$.}
\begin{eqnarray}
\tilde{N}_i & = & \left (m^2_i - \tilde{m}^2_j \right )
\left (m^2_i - \tilde{m}^2_k \right )
+ A \left [\left (m^2_i - \tilde{m}^2_j \right )
|\tilde{V}_{e k}|^2
+ \left (m^2_i - \tilde{m}^2_k \right )
|\tilde{V}_{e j}|^2 \right ] \; ,
\nonumber \\
\tilde{D}^2_i & = & \tilde{N}^2_i + A^2
|\tilde{V}_{e i}|^2 \left [
\left (m^2_i - \tilde{m}^2_j \right )^2
|\tilde{V}_{e k}|^2
+ \left (m^2_i - \tilde{m}^2_k \right )^2
|\tilde{V}_{e j}|^2 \right ] \; .
\end{eqnarray}
Obviously $A=0$ leads to $V = \tilde{V}$. The exact and compact formulas
obtained above show clearly how the flavor
mixing matrix in vacuum is connected to
that in matter. Some instructive analytical approximations can be made
for Eq. (11), once the hierarchy of effective neutrino masses and that of
effective flavor mixing matrix elements are experimentally known.

With the help of Eq. (11), one may calculate the universal
CP-violating parameter $J$ in vacuum.
Indeed it is easier to derive the relationship
between $J$ and $\tilde{J}$ by use of either the sum rules in
Eq. (8) \cite{Xing01} or the equality between
the determinants of lepton mass matrices in matter and
in vacuum \cite{Scott}. The result is
\begin{equation}
J \; = \; \tilde{J} ~
\frac{\Delta \tilde{m}^2_{21}}{\Delta m^2_{21}}\cdot
\frac{\Delta \tilde{m}^2_{31}}{\Delta m^2_{31}}\cdot
\frac{\Delta \tilde{m}^2_{32}}{\Delta m^2_{32}} \;\; ,
\end{equation}
where the neutrino mass-squared differences in vacuum can
be read off from Eq. (9).

Note that the afore-obtained results are valid only
for neutrinos propagating in vacuum and interacting with matter.
As for antineutrinos, the respective formulas for
$m^2_i$, $V$ and $J$ can straightforwardly be
obtained from Eqs. (9) -- (13) through the replacements
$\tilde{V} \Longrightarrow \tilde{V}^*$ and $A \Longrightarrow -A$.

\section{Standard parametrization and numerical illustration}

It is well known that the 3$\times$3 lepton flavor mixing matrix
can be parametrized in terms of three rotation angles
($\theta_1, \theta_2, \theta_3$) and three phase angles
($\delta, \rho, \sigma$), if neutrinos are Majorana particles.
There are nine distinct parametrizations of this
nature \cite{FX01}, but only one of them is particularly convenient in
the analyses of experimental data on the neutrinoless
double beta decay and neutrino oscillations. This ``standard''
parametrization reads in vacuum as
\begin{equation}
V \; = \; \left ( \matrix{
c_1 c_3 & s_1 c_3 & s_3 \cr
- c_1 s_2 s_3 - s_1 c_2 e^{-i\delta} &
- s_1 s_2 s_3 + c_1 c_2 e^{-i\delta} &
s_2 c_3 \cr
- c_1 c_2 s_3 + s_1 s_2 e^{-i\delta} &
- s_1 c_2 s_3 - c_1 s_2 e^{-i\delta} &
c_2 c_3 \cr } \right )
\left ( \matrix{
1	& 0	& 0 \cr
0	& e^{i\rho}	& 0 \cr
0	& 0	& e^{i\sigma} \cr} \right ) \;
\end{equation}
with $s_i \equiv \sin\theta_i$ and $c_i \equiv \cos\theta_i$
(for $i = 1, 2, 3$).
Without loss of generality, the three mixing angles
($\theta_1, \theta_2, \theta_3$) can all be arranged to lie in
the first quadrant. Arbitrary values between $-\pi$ and $+\pi$
are allowed for the CP-violating phases $\delta$,
$\rho$ and $\sigma$. Note that the location of the Dirac-type phase
$\delta$ in $V$ is different from that advocated by the Particle Data
Group in Ref. \cite{PDG}. The advantage of our present phase assignment is
that $\delta$ itself does not appear in the effective Majorana mass
term of the neutrinoless double beta decay \cite{FX01}.
Note also that normal neutrino oscillations are completely insensitive to
the Majorana-type phases $\rho$ and $\sigma$,
therefore the magnitudes of these two parameters keep unchanged even when
neutrinos propagate in matter \cite{Langacker}.
On the other hand, the other four parameters
$\theta_1, \theta_2, \theta_3$ and $\delta$ may be contaminated by
the terrestrial matter effects in realistic long-baseline neutrino
oscillation experiments.
In analogy to Eq. (14), the effective (matter-corrected)
lepton flavor mixing matrix can be parametrized as follows:
\begin{equation}
\tilde{V} \; = \; \left ( \matrix{
\tilde{c}_1 \tilde{c}_3 &
\tilde{s}_1 \tilde{c}_3 &
\tilde{s}_3 \cr
- \tilde{c}_1 \tilde{s}_2 \tilde{s}_3 -
\tilde{s}_1 \tilde{c}_2 e^{-i\tilde{\delta}} &
- \tilde{s}_1 \tilde{s}_2 \tilde{s}_3 +
\tilde{c}_1 \tilde{c}_2 e^{-i\tilde{\delta}} &
\tilde{s}_2 \tilde{c}_3 \cr
- \tilde{c}_1 \tilde{c}_2 \tilde{s}_3 +
\tilde{s}_1 \tilde{s}_2 e^{-i\tilde{\delta}} &
- \tilde{s}_1 \tilde{c}_2 \tilde{s}_3 -
\tilde{c}_1 \tilde{s}_2 e^{-i\tilde{\delta}} &
\tilde{c}_2 \tilde{c}_3 \cr } \right )
\left ( \matrix{
1	& 0	& 0 \cr
0	& e^{i\rho}	& 0 \cr
0	& 0	& e^{i\sigma} \cr} \right ) \;
\end{equation}
with $\tilde{s}_i \equiv \sin\tilde{\theta}_i$ and
$\tilde{c}_i \equiv \cos\tilde{\theta}_i$ (for $i=1, 2, 3$).
The universal CP-violating parameters $J$ and $\tilde J$ defined
in Eq. (2) depend respectively upon the Dirac-type phases $\delta$
and $\tilde \delta$; i.e.,
\begin{eqnarray}
J & = & s_1 c_1 s_2 c_2 s_3 c^2_3 \sin\delta \; ,
\nonumber \\
\tilde{J} & = & \tilde{s}_1 \tilde{c}_1 \tilde{s}_2
\tilde{c}_2 \tilde{s}_3 \tilde{c}^2_3 \sin\tilde{\delta} \; .
\end{eqnarray}
The CP- and T-violating asymmetries of neutrino oscillations
are proportional to $J$ in vacuum and $\tilde J$ in matter.

Now let us proceed to figure out the analytical expressions of the
fundamental parameters
$(\theta_1, \theta_2, \theta_3, \delta)$ in terms of
the effective parameters
$(\tilde{\theta}_1, \tilde{\theta}_2, \tilde{\theta}_3, \tilde{\delta})$,
which can directly be measured from a variety of long-baseline
neutrino oscillation experiments. To do so,
we simply apply the standard parametrization of $\tilde{V}$ to
the parametrization-independent formula in Eq. (11). Then
the matrix elements of $V$ read explicitly as
\begin{equation}
V_{\alpha i} \; =\; \frac{\tilde{N}_i}{\tilde{D}_i}
\tilde{V}_{\alpha i} ~ - ~
\frac{A}{\tilde{D}_i} \sum^3_{k=1} \left (\tilde{T}_{\alpha k}
P_{ki} \right ) \; ,
\end{equation}
where $\tilde{N}_i$ and $\tilde{D}_i$ have been given in Eq. (12),
$P \equiv {\rm Diag} \{1, e^{i\rho}, e^{i\sigma} \}$ denotes
the diagonal Majorana phase matrix, and the matter-associated
quantities $\tilde{T}_{\alpha k}$ are given as
\begin{eqnarray}
\tilde{T}_{e1} & = & + \tilde{c}_1 \tilde{c}_3
\left [ \left ( m^2_1 - \tilde{m}^2_2 \right ) \tilde{s}^2_3
+ \left ( m^2_1 - \tilde{m}^2_3 \right )
\tilde{s}^2_1 \tilde{c}^2_3 \right ] \; ,
\nonumber \\
\tilde{T}_{e2} & = & + \tilde{s}_1 \tilde{c}_3
\left [ \left ( m^2_2 - \tilde{m}^2_1 \right ) \tilde{s}^2_3
+ \left ( m^2_2 - \tilde{m}^2_3 \right )
\tilde{c}^2_1 \tilde{c}^2_3 \right ] \; ,
\nonumber \\
\tilde{T}_{e3} & = & + \tilde{s}_3 \tilde{c}^2_3
\left [ \left ( m^2_3 - \tilde{m}^2_1 \right ) \tilde{s}^2_1
+ \left ( m^2_3 - \tilde{m}^2_2 \right ) \tilde{c}^2_1 \right ] \; ,
\nonumber \\
\tilde{T}_{\mu 1} & = & + \tilde{c}_1 \tilde{c}^2_3
\left [ \left ( m^2_1 - \tilde{m}^2_2 \right ) \tilde{s}_2 \tilde{s}_3
- \left ( m^2_1 - \tilde{m}^2_3 \right )
\left ( \tilde{s}^2_1 \tilde{s}_2 \tilde{s}_3
- \tilde{s}_1 \tilde{c}_1 \tilde{c}_2
e^{-i\tilde{\delta}} \right ) \right ] \; ,
\nonumber \\
\tilde{T}_{\mu 2} & = & + \tilde{s}_1 \tilde{c}^2_3
\left [ \left ( m^2_2 - \tilde{m}^2_1 \right ) \tilde{s}_2 \tilde{s}_3
- \left ( m^2_2 - \tilde{m}^2_3 \right )
\left ( \tilde{c}^2_1 \tilde{s}_2 \tilde{s}_3 +
\tilde{s}_1 \tilde{c}_1 \tilde{c}_2
e^{-i\tilde{\delta}} \right ) \right ] \; ,
\nonumber \\
\tilde{T}_{\mu 3} & = & - \tilde{s}_3 \tilde{c}_3
\left [ \left ( m^2_3 - \tilde{m}^2_1 \right )
\tilde{s}^2_1 \tilde{s}_2 \tilde{s}_3 +
\left ( m^2_3 - \tilde{m}^2_2 \right ) \tilde{c}^2_1 \tilde{s}_2 \tilde{s}_3
- \Delta \tilde{m}^2_{21} \tilde{s}_1 \tilde{c}_1 \tilde{c}_2
e^{-i\tilde{\delta}} \right ] \; ,
\nonumber \\
\tilde{T}_{\tau 1} & = & + \tilde{c}_1 \tilde{c}^2_3
\left [ \left ( m^2_1 - \tilde{m}^2_2 \right ) \tilde{c}_2 \tilde{s}_3
- \left ( m^2_1 - \tilde{m}^2_3 \right )
\left ( \tilde{s}^2_1 \tilde{c}_2 \tilde{s}_3
+ \tilde{s}_1 \tilde{c}_1 \tilde{s}_2
e^{-i\tilde{\delta}} \right ) \right ] \; ,
\nonumber \\
\tilde{T}_{\tau 2} & = & + \tilde{s}_1 \tilde{c}^2_3
\left [ \left ( m^2_2 - \tilde{m}^2_1 \right ) \tilde{c}_2 \tilde{s}_3
- \left ( m^2_2 - \tilde{m}^2_3 \right )
\left ( \tilde{c}^2_1 \tilde{c}_2 \tilde{s}_3
- \tilde{s}_1 \tilde{c}_1 \tilde{s}_2
e^{-i\tilde{\delta}} \right ) \right ] \; ,
\nonumber \\
\tilde{T}_{\tau 3} & = & - \tilde{s}_3 \tilde{c}_3
\left [ \left ( m^2_3 - \tilde{m}^2_1 \right )
\tilde{s}^2_1 \tilde{c}_2 \tilde{s}_3 +
\left ( m^2_3 - \tilde{m}^2_2 \right ) \tilde{c}^2_1 \tilde{c}_2 \tilde{s}_3
- \Delta \tilde{m}^2_{21} \tilde{s}_1 \tilde{c}_1 \tilde{s}_2
e^{-i\tilde{\delta}} \right ] \; .
\end{eqnarray}
It should be noted that the matrix elements $V_{\mu 3}$ and $V_{\tau 3}$
in Eq. (17) are complex and dependent on the effective
CP-violating phase $\tilde{\delta}$, as one can see from
$\tilde{T}_{\mu 3}$ and $\tilde{T}_{\tau 3}$ in Eq. (18).
Hence a proper redefinition of the phases for muon and tau fields is
needed, in order to make $V_{\mu 3}$ and $V_{\tau 3}$ real in the course
of linking $\tilde{V}$ in Eq. (15) to $V$ in Eq. (14).
The instructive relations between the effective mixing angles
in matter ($\tilde{\theta}_1, \tilde{\theta}_2, \tilde{\theta}_3$)
and the fundamental mixing angles in vacuum
($\theta_1, \theta_2, \theta_3$) are found to be
\begin{eqnarray}
\frac{\tan\theta_1}{\tan\tilde{\theta}_1} & = &
\frac{\displaystyle
\tilde{N}_2 - A \left [ \left ( m^2_2 - \tilde{m}^2_1 \right )
\tilde{s}^2_3 +
\left ( m^2_2 - \tilde{m}^2_3 \right )
\tilde{c}^2_1 \tilde{c}^2_3 \right ] }
{\displaystyle
\tilde{N}_1 - A \left [ \left ( m^2_1 - \tilde{m}^2_2 \right )
\tilde{s}^2_3 +
\left ( m^2_1 - \tilde{m}^2_3 \right ) \tilde{s}^2_1 \tilde{c}^2_3 \right ] }
\cdot \frac{\tilde{D}_1}{\tilde{D}_2} \; ,
\nonumber \\
\frac{\tan\theta_2}{\tan\tilde{\theta}_2} & = &
1 - \frac{ A \Delta \tilde{m}^2_{21} \tilde{s}_1 \tilde{c}_1
\tilde{s}_3 \cos\tilde{\delta}/\left ( \tilde{s}_2 \tilde{c}_2 \right )}
{\displaystyle
\tilde{N}_3 + A \left [ \left ( m^2_3 - \tilde{m}^2_1 \right )
\tilde{s}^2_1 +
\left ( m^2_3 - \tilde{m}^2_2 \right ) \tilde{c}^2_1 \right ]
\tilde{s}^2_3 } \; ,
\nonumber \\
\frac{\sin\theta_3}{\sin\tilde{\theta}_3} & = &
\frac{\tilde{N}_3}{\tilde{D}_3} ~ - ~
\frac{A}{\tilde{D}_3} \left [ \left ( m^2_3 - \tilde{m}^2_1 \right )
\tilde{s}^2_1 + \left ( m^2_3 - \tilde{m}^2_2 \right )
\tilde{c}^2_1 \right ] \tilde{c}^2_3 \; .
\end{eqnarray}
Note that we have only presented the next-to-leading order expression
for $\tan\theta_2/\tan\tilde{\theta}_2$ in Eq. (19), since the exact result
is too complicated to be instructive. The former works to a
high degree of accuracy, and it is identical to the exact result provided
that $\tilde{\theta}_2 = \pi/4$ and $\tilde{\delta} = \pm \pi/2$ hold.
Once the relations between $\tilde{\theta}_i$ and $\theta_i$ (for
$i=1, 2, 3$) are fixed, one can derive the relation between the
effective CP-violating phase $\tilde{\delta}$ in matter and the
genuine CP-violating phase $\delta$ in vacuum
by use of Eq. (16) as well as
the relationship between $\tilde{J}$ and $J$ in Eq. (13). We obtain
\begin{equation}
\frac{\sin\delta}{\sin\tilde{\delta}} \; =\;
\frac{\tilde{s}_1 \tilde{c}_1 \tilde{s}_2 \tilde{c}_2
\tilde{s}_3 \tilde{c}^2_3}{s_1 c_1 s_2 c_2 s_3 c^2_3}
\cdot \frac{\Delta \tilde{m}^2_{21}}{\Delta m^2_{21}}
\cdot \frac{\Delta \tilde{m}^2_{31}}{\Delta m^2_{31}}
\cdot \frac{\Delta \tilde{m}^2_{32}}{\Delta m^2_{32}} \;\; .
\end{equation}
Of course, $\tilde{\delta} = \delta$ holds in the limit $A = 0$. The
formulas obtained in Eqs. (19) and (20) are very useful for the purpose
of extracting the fundamental parameters of lepton flavor mixing from the
matter-corrected ones, which can be measured from various long-
and medium-baseline neutrino oscillation experiments in the near future.

Note again that the afore-obtained results are valid only
for neutrinos interacting with matter.
As for antineutrinos, the corresponding expressions of
$\theta_1$, $\theta_2$, $\theta_3$ and $\delta$ can straightforwardly be
achieved from Eqs. (19) and (20) through the replacements
$\tilde{\delta} \Longrightarrow -\tilde{\delta}$ and $A \Longrightarrow -A$.

It is worthwhile at this point to remark two advantages of the results
presented above over those obtained by Zaglauer and
Schwarzer in Ref. \cite{Zaglauer}. First, our relations clearly show the
proportionality between the sine or tangent functions of
$(\theta_1, \theta_2, \theta_3, \delta)$ and
$(\tilde{\theta}_1, \tilde{\theta}_2, \tilde{\theta}_3, \tilde{\delta})$.
Second, our relations make the dependence of
$\tilde{\theta}_1$, $\tilde{\theta}_2$, $\tilde{\theta}_3$ and
$\tilde{\delta}$ on the matter parameter $A$ more transparent.
Therefore our results are expected to be more useful for the analytical
study of terrestrial matter effects on lepton flavor mixing and CP violation.

For numerical illustration, we assume the matter density of the earth's crust
to be constant. Then $A \approx 2.28 \cdot 10^{-4} ~ {\rm eV}^2 E/[{\rm GeV}]$
is a good approximation \cite{M},
where $E$ is the neutrino beam energy. We typically
take $E =5$ GeV. The input values of the effective neutrino mixing parameters
in matter are listed on the left side of Table 1.
With the help of Eqs. (9) and (10) as well
as Eqs. (19) and (20), one may calculate the fundamental neutrino mixing
parameters in vacuum. We list our numerical results on the right side of
Table 1 for both the case of neutrinos ($+A$) and that of
antineutrinos ($-A$). One can see that the input values of
$(\Delta \tilde{m}^2_{21}, \Delta \tilde{m}^2_{31})$ and
$(\tilde{\theta}_1, \tilde{\theta}_2, \tilde{\theta}_3, \tilde{\delta})$
are rather {\it ad hoc}. The reason is simply that we contrive to
get the {\it appropriate} sizes of $(\Delta m^2_{21}, \Delta m^2_{31})$ and
$(\theta_1, \theta_2, \theta_3, \delta)$, which are compatible with
the present Super-Kamiokande data on solar and atmospheric neutrino
oscillations \cite{SK}. Indeed we observe from Table 1 that
$\Delta m^2_{\rm sun} \approx \Delta m^2_{21}
\approx 4.62 \cdot 10^{-5} ~ {\rm eV}^2$
(or $5.05 \cdot 10^{-5} ~ {\rm eV}^2$)
and  $\sin^2 2\theta_{\rm sun} \approx \sin^2 2\theta_1 \approx 0.75$
(or $0.90$) do agree with the large-angle MSW solution to the solar
neutrino problem, and $\Delta m^2_{\rm atm} \approx \Delta m^2_{31}
\approx 3.00 \cdot 10^{-3} ~ {\rm eV}^2$
(or $2.89 \cdot 10^{-3} ~ {\rm eV}^2$) together with
$\sin^2 2\theta_{\rm atm} \approx \sin^2 2\theta_{2} \approx 0.99$
(or $0.97$) is consistent with the atmospheric neutrino oscillation data.
Moreover, $\sin^2 2\theta_{\rm CHOOZ} \approx
\sin^2 2\theta_3 \approx 0.03 < 0.1$ coincides with the CHOOZ and
Palo Verde reactor experiments \cite{CHOOZ}.
Note that $\Delta \tilde{m}^2_{21} \sim A \gg \Delta m^2_{21}$ holds
for the given value of $E$, hence there appears a substantial difference
between $\theta_1$ and $\tilde{\theta}_1$. On the other hand,
$\theta_2 \approx \tilde{\theta}_2$ and $\delta \approx \tilde{\delta}$
hold to an excellent degree of accuracy.
The tiny difference between $\theta_2$ and $\tilde{\theta}_2$ in our
numerical illustration can easily be understood: the special input
$\tilde{\delta} = \pm 90^\circ$ assures the vanishing of the
term proportional to $A \cos \tilde{\delta}$ on the right-hand side
of the expression for $\tan\theta_2/\tan\tilde{\theta}_2$ in Eq. (19),
leading straightforwardly to $\theta_2 = \tilde{\theta}_2$ in the
next-to-leading order approximation. In general,
$\theta_2 \approx \tilde{\theta}_2$ remains a good approximation,
if the condition $|\Delta m^2_{21}| \ll |\Delta m^2_{31}|$ is
satisfied \cite{Barger00}. One may numerically check that
$\delta \approx \tilde{\delta}$ is generally valid as well, provided that
$|\Delta m^2_{21}| \ll |\Delta m^2_{31}|$ holds. Analytically, however,
we have not found a transparent way to show the approximate equality
between $\delta$ and $\tilde{\delta}$. Some more effort is obviously
desirable, in order to understand why the CP-violating phase $\delta$ is
in most cases insensitive to the terrestrial matter effects \cite{Xing}.

Our numerical results illustrate
that the fundamental neutrino mixing parameters can straightforwardly be
extracted from their matter-corrected counterparts, once the latter are
measured from the future long-baseline neutrino oscillation experiments.
In particular, $|\Delta m^2_{21}| \ll |\Delta m^2_{31}|$ assures that
$\tilde{\theta}_2 \approx \theta_2$ and $\tilde{\delta} \approx \delta$
hold in the standard parametrization of lepton flavor mixing.

\section{Summary}

We have pointed out that there exists the reversibility between the
fundamental neutrino mixing parameters in vacuum and their effective
counterparts in matter. The former can therefore be expressed in terms
of the latter, allowing more straightforward extraction of the genuine
lepton mixing quantities from a variety of long-baseline neutrino
oscillation experiments. Besides the parametrization-independent
formalism, we have presented the formulas based on the standard
parametrization of the neutrino mixing matrix and given a typical numerical
illustration. Our results are expected to be a useful addition to
the phenomenology of lepton flavor mixing and neutrino oscillations.

\newpage

\begin{table}[t]
\caption{Numerical illustration of extracting the fundamental neutrino mixing
parameters in vacuum from their effective counterparts in matter,
where $A = 2.28 \cdot 10^{-4} ~ {\rm eV}^2 E/[{\rm GeV}]$ with
$E = 5$ GeV has typically been used.}
\vspace{0.1cm}
\begin{center}
\begin{tabular}{lcl} \hline
& Neutrinos ($+A$) \\
Input Parameters &~~~~~~~~~~~~~~~~& Output Parameters  \\ \hline
$\Delta \tilde{m}^2_{21} = 1.15 \cdot 10^{-3} ~ {\rm eV}^2$
&& $\Delta m^2_{21} = 4.62 \cdot 10^{-5} ~ {\rm eV}^2$ \\
$\Delta \tilde{m}^2_{31} = 3.00 \cdot 10^{-3} ~ {\rm eV}^2$
&& $\Delta m^2_{31} = 3.00 \cdot 10^{-3} ~ {\rm eV}^2$ \\
$\tilde{\theta}_1 = 89^\circ$
&& $\theta_1 = 60^\circ$ \\
$\tilde{\theta}_2 = 42^\circ$
&& $\theta_2 = 42^\circ$ \\
$\tilde{\theta}_3 = 8^\circ$
&& $\theta_3 = 5^\circ$ \\
$\tilde{\delta} = 90^\circ$
&& $\delta = 90^\circ$ \\ \hline \hline
& Antineutrinos ($-A$) \\
Input Parameters &~~~~~~~~~~~~~~~~& Output Parameters  \\ \hline
$\Delta \tilde{m}^2_{21} = 1.15 \cdot 10^{-3} ~ {\rm eV}^2$
&& $\Delta m^2_{21} = 5.05 \cdot 10^{-5} ~ {\rm eV}^2$ \\
$\Delta \tilde{m}^2_{31} = 4.00 \cdot 10^{-3} ~ {\rm eV}^2$
&& $\Delta m^2_{31} = 2.89 \cdot 10^{-3} ~ {\rm eV}^2$ \\
$\tilde{\theta}_1 = 1.2^\circ$
&& $\theta_1 = 36^\circ$ \\
$\tilde{\theta}_2 = 40^\circ$
&& $\theta_2 = 40^\circ$ \\
$\tilde{\theta}_3 = 3.6^\circ$
&& $\theta_3 = 5^\circ$ \\
$\tilde{\delta} = -90^\circ$
&& $\delta = -90^\circ$ \\ \hline
\end{tabular}
\end{center}
\end{table}
\normalsize

\end{document}